# Discovery of a binary icosahedral quasicrystal in $Sc_{12}Zn_{88}$

P.C. Canfield[1,2*], M.L. Caudle[2], C.-S. Ho[1†], A. Kreyssig[1,2], S. Nandi[1,2], M. G. Kim[1,2], X. Lin[1,2], A. Kracher[1], K. W. Dennis[1], R.W. McCallum[1,3] and A. I. Goldman[1,2*]

[1]Ames Laboratory, US DOE, Iowa State University, Ames, Iowa 50011
[2]Department of Physics and Astronomy, Iowa State University, Ames, Iowa 50011
[3]Department of Materials Science and Engineering, Iowa State University, Ames, Iowa 50011

[*]email: goldman@ameslab.gov, canfield@ameslab.gov

**Abstract**

We report the discovery of a new binary icosahedral phase in a Sc-Zn alloy obtained through solution-growth, producing millimeter-sized, facetted, single grain, quasicrystals that exhibit different growth morphologies, pentagonal dodecahedra and rhombic triacontahedra, under only marginally different growth conditions. These two morphologies manifest different degrees of quasicrystalline order, or phason strain. The discovery of i-$Sc_{12}Zn_{88}$ suggests that a reexamination of binary phase diagrams at compositions close to crystalline approximant structures may reveal other, new binary quasicrystalline phases.

Icosahedral quasicrystals, discovered 25 years ago[1], possess long-range positional and orientational order, but lack the periodic translational order of crystalline solids[2]. Although they form, almost readily, in a wide variety of ternary and quaternary metallic alloys, examples of stable binary icosahedral quasicrystals are quite rare. Indeed, it has been nearly a decade since the discovery of the only known stable binary icosahedral phases in Cd-Yb and Cd-Ca[3,4]. Binary alloys offer several important advantages, not the least of which is their simplicity, relative to ternary and quaternary compositions, for structural studies and theoretical treatments. The Sc-Zn system has been associated with quasicrystal formation for some time[5].

An icosahedral phase has been identified in $Sc_{15}Mg_5Zn_{80}$[6] and a large number of ternary and quaternary compositions of the type Sc-M-Zn (M = Mn, Fe, Co, Ni, Cu, Pd, Pt, Au, Ag). In addition, the binary crystalline alloy, $ScZn_6$ (formerly identified as $Sc_3Zn_{17}$)[7], has been identified as a approximant of the icosahedral phase. Despite all of this study, the binary "parent" icosahedral phase in the Sc-Zn system has eluded detection. It was then startling when, as part of the exploration of the Sc-Zn binary phase diagram[8], we discovered millimeter-sized facetted grains, with icosahedral morphologies, in flux-grown samples.

Single grains of $Sc_{12}Zn_{88}$ quasicrystals were produced using traditional solution growth methods[9,10]. Starting compositions, ranging from $Sc_4Zn_{96}$ to $Sc_2Zn_{98}$, were heated to 950°C, cooled to 800°C over several hours and then slowly cooled to 480°C over 30-70 hours. At this temperature the crucibles were removed from the furnace and the remaining liquid decanted. Initial growths were performed using two 2 milliliter alumina crucibles, one inverted over the other, with silica wool acting as a filtering media during decanting[9]. Given the substantial vapor pressure of Zn, later growths utilized specially designed, threaded alumina crucibles with a threaded alumina straining layer, similar in function to the Ta crucibles described in reference 10. For all growths, high purity Sc and Zn were placed in the lower crucible and both crucibles were sealed in a silica tube back filled with approximately 0.25 atmospheres of high purity Ar. Upon opening the crucibles small facetted grains (Fig. 1) in the shape of either a pentagonal dodecahedron (PD) or rhombic triacontahedron (RT) were found either on the surfaces of crystals of the cubic $ScZn_6$ phase, on the strainer surface, or on the walls of the crucible itself.

The PD facetted grains were most commonly grown from initial compositions of $Sc_4Zn_{96}$ and $Sc_3Zn_{97}$ and found in the presence of well formed and faceted, cubic $ScZn_6$. The RT facetted grains were more commonly grown from initial compositions of $Sc_2Zn_{98}$ and also found in the presence of $ScZn_6$ but, due to the reduced Sc content of the initial melt, less of the cubic phase was in evidence. If the growths were cooled to temperatures lower than 480°C, then plate-like crystals of the tetragonal $ScZn_{12}$ phase[11] also formed.

Wavelength Dispersive X-ray Spectroscopy was used to establish the stoichiometry of the PD and RT grains and to rule out the possibility of a third element impurity, leading to a ternary phase. The composition of both the PD and RT grains was found to be a Sc-Zn binary with 12 ±0.3%-atomic Sc, and no detectable third elements. Differential

Scanning Calorimetry (DSC) was employed to determine the decomposition temperature of this new phase (505°C) by heating samples of isolated pentagonal grains, as well as masses of RT grains (see inset to Figure 2). Taken together, these measurements provided a basis for the revised binary phase diagram in the zinc-rich region presented in Figure 2[12]. A full re-evaluation of the phase diagram as well as possible rate and nucleation dependencies of the QC phase is in order, and ongoing.

The structural characterization of these single grains was accomplished through conventional x-ray powder diffraction measurements and single grain, high-energy x-ray diffraction, shown in Fig. 3. Turning first to the growths that produce PD facets, several grains were crushed and prepared for the powder diffraction measurement. All diffraction peaks could be indexed, using the Elser scheme[13], by a primitive (P-type) icosahedral phase pattern. Using the strongest peak along the fivefold axis (labeled 211111 in Fig. 3a) the quasilattice constant is $a_R = 5.017(3)$ Å.

Both i-$Sc_{12}Zn_{88}$ ($ScZn_{7.3}$) and i-$YbCd_{5.7}$ are close in composition to the 1/1 crystalline approximants, $ScZn_6$[7] and $YbCd_6$[14], although the icosahedral alloy compositions are found on different (e.g. Zn-rich and Zn-poor) sides of the 1:6 composition of their respective approximants[12,15]. Approximants[16] are large unit cell periodic structures closely related to quasicrystals (QC) in that they share the same basic atomic motifs, or clusters, arranged periodically in the approximant (e.g. as a body-centered cubic packing in $ScZn_6$ and $YbCd_6$). Indeed, crystalline approximants have been used to great advantage for investigations of the structural and electronic properties of their quasicrystalline counterparts[17,18,19]. We can compare our measured value for $a_R$ to the lattice constant expected for a 1/1 cubic approximant phase through the relation:

$$a_{\frac{q}{p}} = \frac{2a_R(p+q\tau)}{\sqrt{(2+\tau)}}, \text{ where } \frac{q}{p} \text{ denotes the order of the approximant (1/1 in this case) and}$$

$\tau$ is the golden mean ($\frac{\sqrt{5}+1}{2}$). We find a calculated value of $a_{1/1} = 13.811(8)$ Å. The lattice parameter for $ScZn_6$ (space group $Im3$) is reported as $a_{1/1} = 13.8311(5)$ Å [7]. This close agreement argues that $ScZn_6$ will provide a good starting point for the structural refinement of i-$Sc_{12}Zn_{88}$. Moreover, we note that $ScZn_6$ is isostructural with $YbCd_6$, so that both i-$Sc_{12}Zn_{88}$ and i-$YbCd_{5.7}$ may be described by similar atomic decoration schemes[20], even though there are significant differences between the $\frac{Sc}{Zn}$ and $\frac{Yb}{Cd}$ ratios. Within the context of a Hume-Rothery mechanism for the stabilization of the quasicrystals[2], i-$Sc_{12}Zn_{88}$ yields an e/a (electrons/atom) ratio of 2.12, well within the range of 2.01-2.16 reported for Sc-M-Zn icosahedral alloys and 2.00 for i-$YbCd_{5.7}$[21].

In Figs. 3c and 3e we show the results of high-energy x-ray diffraction measurements taken on station 6ID-D in the MUCAT Sector at the Advanced Photon Source with the incident beam along the five-fold (Fig. 3c) and two-fold (Fig. 3e) axes of a single PD facetted grain. These data were taken using an incident x-ray energy of 129 keV and recorded on a MAR 345 area detector. The use of high energy x-rays ensures that the structure of the sample bulk (rather than surface) is probed and, by rocking the sample through small angular ranges about axes perpendicular to the beam, provides an image of reciprocal space planes that lie normal to the beam direction[22]. All diffraction spots in the twofold and fivefold planes can be indexed to the icosahedral phase and lie, within the resolution of the detector, at the predicted positions. Note that as one moves from the center to the periphery of the pattern, diffraction spots from higher order zones are also in evidence. For instance, the weak peak in the center of pentagon B in Fig. 3c is not observed in neighboring pentagon A since it does not originate in the five-fold plane, but slightly above it (within the range over which the sample is rocked). Close inspection of the sequence of diffraction peaks along the fivefold axis in Figure 3c does not reveal any peaks at positions expected for a face-centered icosahedral quasilattice, confirming that the structure is P-type icosahedral. Furthermore, the high energy x-ray diffraction patterns establish that the full 0.5 mm$^3$ volume of the sample shares the same orientation, so that the grain is properly characterized as a "single quasicrystal." However, the longitudinal widths of diffraction peaks measured along the twofold and fivefold axes are not resolution limited and we find a systematic broadening consistent with the presence of "phason strain", a type of disorder characteristic of primitive icosahedral quasilattices[2]. This is particularly interesting in light of the high degree of quasicrystalline order noted for some of the P-type ternary Sc-Mg-Zn icosahedral alloys[21] where the authors noted that the formation of the icosahedral phase was quite sensitive to the Mg concentration.

The fact that the PD grains are bulk quasicrystals, taken together with the method of growth, provides strong evidence that i-$Sc_{12}Zn_{88}$ is a stable phase. The quasicrystal forms during a 30-70 hour cooling process to 480°C followed by cooling to room temperature over approximately 10 minutes. We have also annealed as-grown samples at 390°C for 22 hours in one atmosphere of Ar and found no or morphological or structural changes as evidenced by visual inspection, the x-ray Laue pattern, and x-ray powder diffraction. If i-$Sc_{12}Zn_{88}$ is not thermodynamically stable at room temperature, it is quite robust to the temperature-time treatment described above to the degree of remaining a bulk single grain quasicrystal as evidenced by Figs. 2 and 3.

For the RT facetted grains, Figs. 3b, d and f show the powder and single grain high energy x-ray diffraction measurements analogous to those for the PD sample. The powder diffraction profile for the RT sample is shown Fig. 3b and yields the same value for $a_R$, as the PD sample. A comparison between features observed in the high-energy x-ray diffraction measurements, however, show some interesting differences. A distinct smearing of the diffraction peaks (on the order of one degree) in the transverse direction was observed for all of the RT grains investigated and the longitudinal (radial) diffraction peak widths are approximately a factor of two broader than those of the PD grains. The peak positions remain well-indexed by the icosahedral phase but the peak broadening points to a lower degree of "quasicrystallinity" in the RT grains, relative to the PD grains. The finite mosaic of these samples may arise from misorientations between different regions of the grain that may be related to the growth conditions for the RT grains. The differences in longitudinal widths are consistent with a larger degree of phason strain in the RT grains, a point that deserves further detailed study.

i-Sc$_{12}$Zn$_{88}$ offers new opportunities for investigations of the structure, formation, growth and stability of aperiodic alloys.  For example, this new binary icosahedral phase exhibits different growth morphologies under only marginally different growth conditions, and we have found that the degree of quasicrystalline order differs between the PD and RT grains.  Together with the ternary Sc-M-Zn icosahedral alloys, i-Sc$_{12}$Zn$_{88}$ offers a new avenue for systematic investigations of the effects of chemical substitution as well as formation and growth conditions on quasicrystalline order.  Furthermore, the x-ray cross-sections for Sc (Z=21) and Zn (Z=30) provide reasonable contrast for structural measurements but, relative to i-YbCd$_{5.7}$,  i-Sc$_{12}$Zn$_{88}$ also offers the advantage of allowing both neutron diffraction and inelastic scattering measurements on samples without the need for isotopic substitution (Cd is highly neutron absorbing).

Finally, it is worth noting that the proposed region of exposed liquidus line for primary solidification of the Sc$_{12}$Zn$_{88}$ icosahedral phase is very limited (Fig. 2).  It is not surprising, then, that this phase eluded discovery for so long, despite recent studies of related binary approximant phases.  This said, the obvious and tantalizing question is how many other quasicrystalline phases have been missed (or are yet to be discovered) in the proximity of binary approximants?  Solution growth experiments, that sample the cascading peritectics associated with binary liquidus lines, may lead to new, as yet undiscovered binary quasicrystalline alloys.

Acknowledgments

The authors gratefully acknowledge the assistance of Douglas Robinson with the high energy x-ray diffraction measurements and Thomas Lograsso, Matthew Kramer and Patricia Thiel for their assistance and useful discussions. We acknowledge the Ames Laboratory Materials Preparation Center for the elemental scandium used in this study.  Work at the Ames Laboratory was supported by the U.S. Department of Energy, Basic Energy Sciences under Contract No. DE-AC02-07CH11358.  The use of the Advanced Photon Source was supported by the U.S. DOE under Contract No. DE-AC02-06CH11357.

† Current address: Department of Physics University of California, Berkeley

[1] D. Shechtman, I. Blech, D. Gratias, and J.W. Cahn, *Phys. Rev. Lett.* **53**, 1951 (1984).

[2] C. Janot, *Quasicrystals: A Primer* (Oxford Univ. Press, New York, 1992).

[3] A.P. Tsai, J.Q. Guo, E. Abe, H. Takakura and T.J. Sato, *Nature* **408**, 537 (2000).

[4] J.Q. Guo, E. Abe, and A.P. Tsai, *Phys. Rev. B* **62**, R14605 (2000).

[5] W. Steurer and S. Deloudi, *Acta Cryst.* A**64**, 1 (2007).

[6] Y. Kaneko, Y. Arichika and T. Ishimasa, *Phil. Mag. Lett.* **81**, 777 (2001).

[7] Q. Lin and J.D. Corbett, *Inorg. Chem.* **43**, 1912 (2004).

[8] P.C. Canfield, P.C., *Nature Physics* **4**, 167 (2008).

[9] P.C. Canfield and Z. Fisk, *Phil. Mag. B*, **65,** 1117 (1992).

[10] P.C. Canfield and I.R. Fisher, *Journal of Crystal Growth* **225**, 155 (2001).

[11] P.I. Kripyakevich, V.S. Protasov and Yu. B.Kuz'ma, *Inorg. Mater. (Engl. Transl.)* **2** 1351 (1966).

[12] A. Palenzona and P. Manfrinetti, *J. Alloys and Compounds* **247**, 195 (1997).

[13] V. Elser, *Phys. Rev. B* **32**, 4892-4898 (1985).

[14] C.P. Gómez and S. Lidin, *Angew. Chem., Int.* **40**, 4037 (2001).

[15] A. Palenzona, *J. Less-Common Metals* **25**, 367 (1971).

[16] See, for example, A.I. Goldman and K,F, Kelton, *Rev. Mod. Phys.* **65**, 213 (1993).

[17] H. Takakura, C.P. Gómez, A. Yamamoto, M. deBoissieu and A.P. Tsai, *Nature Mater.* **6**, 58 (2007).

[18] M. De Boissieu, S. Francoual, M. Mihalovič, K. Shibata, A.Q.R. Baron, Y. Sidis, T. Ishimasa, D. Wu, T. Lograsso, L.-P. Regnault, F. Gähler, S. Tsutsui, B. Hennion, P. Bastie, T.J. Sato, H. Takakura, R. Currat, R. and A.P. Tsai, *Nature Mater.* **6**, 977 (2007).

[19] J. Hasegawa, R. Tamura and S. Takeuchi, *Phys. Rev. B* **66**, 132201 (2002).

[20] T. Ishimasa, Y. Kaneko and H. Kaneko, *J. Alloys and Compounds* **342**, 13 (2002).

[21] T. Ishimasa, Y. Kaneko and H. Kaneko, *J. Non-Cryst. Solids* **334&335**, 1 (2004).

[22] A. Kreyssig, S. Chang, Y. Janssen, J.W. Kim, S. Nandi, J.Q. Yan, L. Tan, R.J. McQueeney, P.C. Canfield and A.I. Goldman, *Phys. Rev. B* **76**, 054421 (2007).

**FIGURE CAPTIONS**

**Figure 1. (color online)** Growth morphologies of solution-grown i-$Sc_{12}Zn_{88}$. Panel **a** shows an example of the pentagonal dodecahedral (PD). Panel **b** shows a cluster of facetted grains displaying a rhombic triacontahedral (RT) morphology. The inset shows an isolated RT grain.

**Figure 2. (color online)** Revised Sc-Zn phase diagram of reference. The Sc-Zn binary phase diagram taken from reference 12 is modified to include the quasicrystalline phase (labelled **Q**). The inset presents the DSC data on the thermal decomposition of the quasicrystalline phase at 505°C.

**Figure 3. (color online)** Powder diffraction and high energy x-ray diffraction from single grains of i-$Sc_{12}Zn_{88}$. Powder diffraction pattern from powders of the PD facetted grains and RT facetted grains are shown in panels **a** and **b**, respectively. Icosahedral indices using Elser's notation[13] are shown for the strongest peaks. Single grain high energy x-ray diffraction patterns measured along the fivefold axis and twofold axis are shown in panels **c** and **e** for the PD grain and **d** and **f** for the RT sample. The lines in panel **e** indicate the twofold, threefold and fivefold directions in the twofold scattering plane. The weak peak in the center of pentagon B in panel **c** is not observed in neighboring pentagon A since it does not originate in the five-fold plane, but slightly above it (within the range over which the sample is rocked).

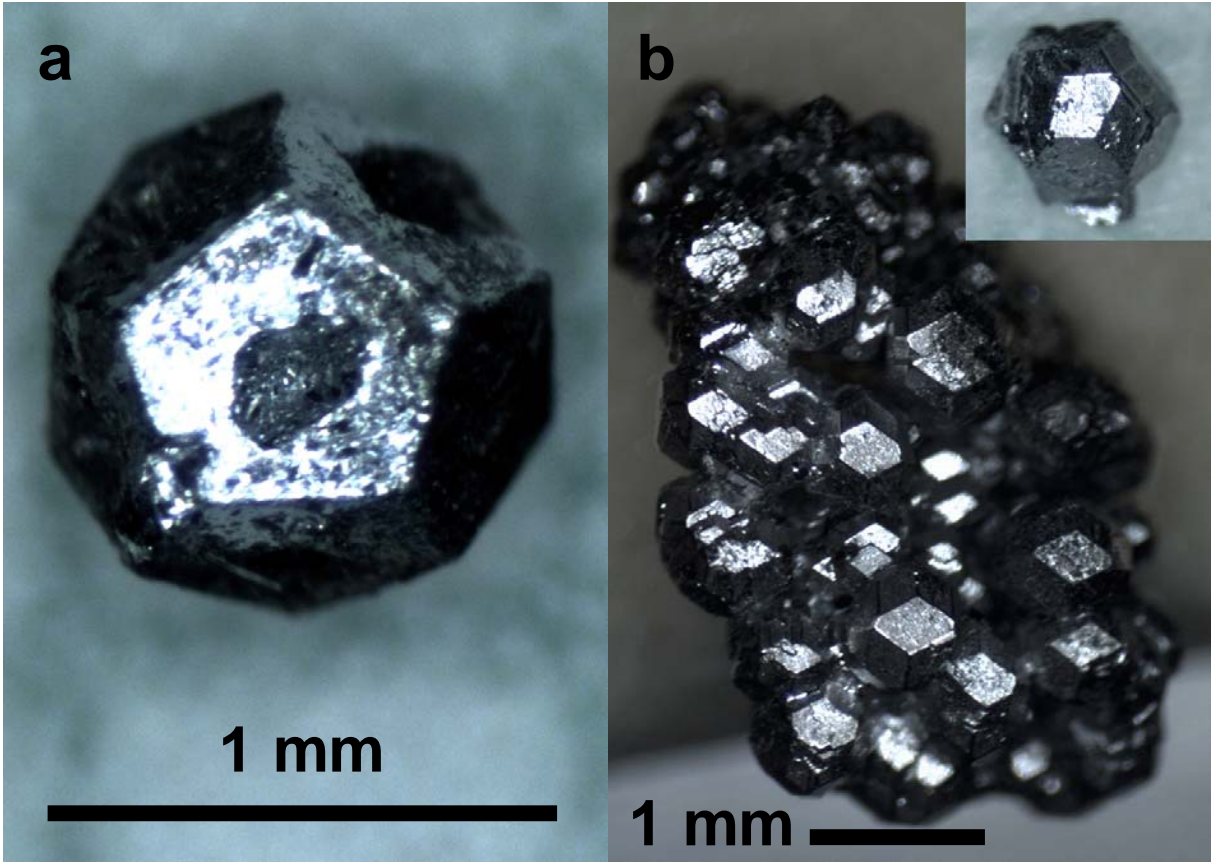

**FIGURE 1**

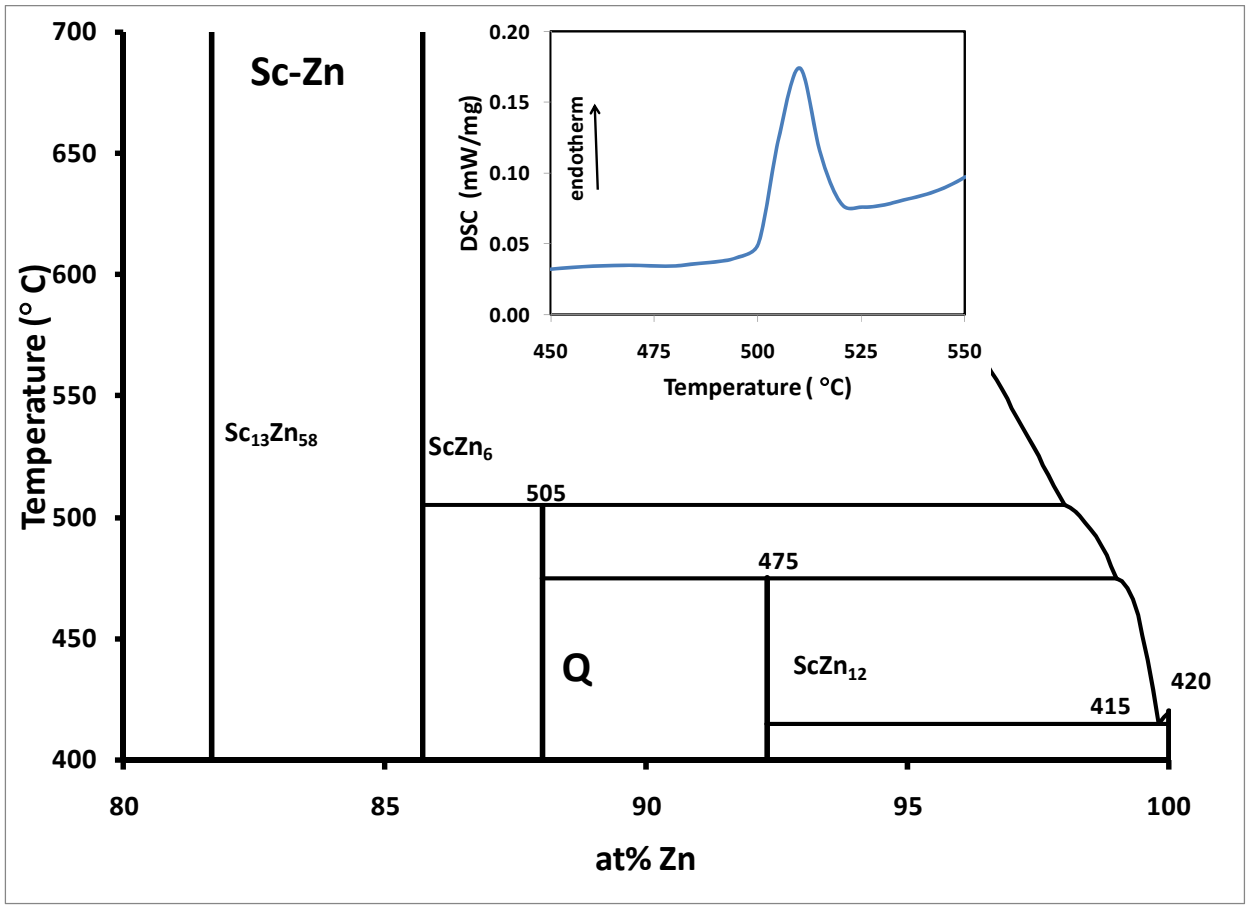

**FIGURE 2**

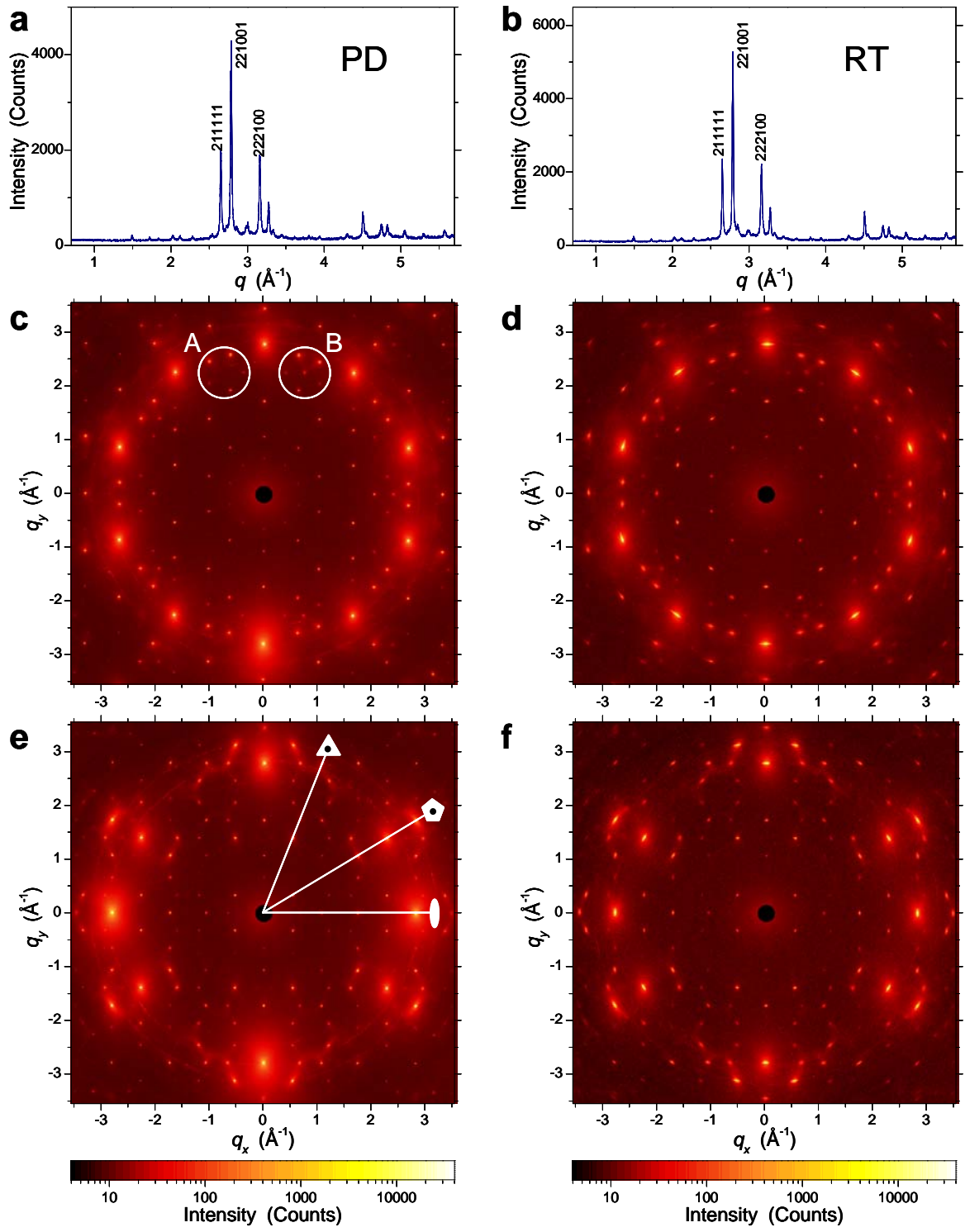

**FIGURE 3**